\documentclass[aps,pra,twocolumn,showpacs,superscriptaddress,amssymb]{revtex4-1}

\usepackage{graphicx}
\usepackage{dcolumn}
\usepackage{bm}
\usepackage{amsmath} 
\graphicspath{{pict/}{}}

\newcommand\pictc[5]{\begin{figure}
                      \centerline{
\includegraphics[width=#1\columnwidth,height=0.7\textheight,keepaspectratio]{#3}}
                      \protect\caption{\protect\label{fig:#4} #5}
                   \end{figure}            }

\newcommand\pict[4][1]{\pictc{#1}{!tb}{#2}{#3}{#4}}

\newcommand\rpict[1]{\ref{fig:#1}}

\newcommand\leqt[1]{\protect\label{eq:#1}}
\newcommand\reqtn[1]{\ref{eq:#1}}
\newcommand\reqt[1]{(\reqtn{#1})}

\newcommand\lsect[1]{\protect\label{sect:#1}}
\newcommand\rsect[1]{\ref{sect:#1}}

\begin{document}

\title{Classical Simulation of Squeezed Vacuum in Optical Waveguide Arrays}

\author{Andrey A. Sukhorukov}
\author{Alexander S. Solntsev}
\affiliation{Nonlinear Physics Centre, Research School of Physics and Engineering, The Australian National University, Canberra, Australia}

\author{John Sipe}
\affiliation{Department of Physics and Institute for Optical Sciences, University of Toronto, 60 St. George St., Toronto, Ontario, Canada}

\begin{abstract}
We reveal that classical light diffraction in arrays of specially modulated coupled optical waveguides can simulate the quantum process of two-mode squeezing in nonlinear media, with the waveguide mode amplitudes corresponding the signal and idler photon numbers.
The whole Fock space is mapped by a set of arrays, where each array represents the states with a fixed difference between the signal and idler photon numbers.
We demonstrate a critical transition from photon number growth to Bloch oscillations with periodical revivals of an arbitrary input state, associated with an increase of the effective phase mismatch between the pump and the squeezed photons.
\end{abstract}


\maketitle

\section{Introduction}  \lsect{introduction}

Arrays of coupled optical waveguides enable flexible manipulation of optical beams~\cite{Christodoulides:2003-817:NAT, Lederer:2008-1:PRP}. Remarkably, the beam evolution can emulate the fundamental wave phenomena in different physical systems~\cite{Garanovich:2012-1:PRP}, including the famous examples of Bloch oscillations~\cite{Peschel:1998-1701:OL, Pertsch:1999-4752:PRL, Morandotti:1999-4756:PRL}, Zener tunneling~\cite{Trompeter:2006-53903:PRL}, and dynamic localization~\cite{Lenz:2003-87:OC, Longhi:2006-243901:PRL, Szameit:2009-271:NPHYS}
originally predicted for electrons in crystalline potentials. Recently, it was shown that classical beam propagation in linear arrays can simulate quantum photonics phenomena including quantum walks~\cite{Perets:2008-170506:PRL}, transformation of quantum coherent and displaced Fock states~\cite{Perez-Leija:2010-2409:OL, Keil:2011-103601:PRL, Perez-Leija:2012-13848:PRA, Keil:2012-3801:OL}, and generation of photon pairs through spontaneous parametric down-conversion~\cite{Grafe:2012-562:SRP}. These analogies on the one hand enable direct visualization of the simulated photon states, and on the other hand suggest novel concepts for the design of waveguide arrays offering new opportunities for the manipulation of optical beams.

In this paper, we develop a special design of waveguide arrays, in which linear light propagation can simulate the dynamics of squeezing. Optical squeezing is a fundamentally important quantum process, which can occur in media with quadratic nonlinearity. Squeezed light can enable high-resolution measurements, beating the standard quantum limit~\cite{*[{}] [{, Sec. 7.7.}]  Gerry:2004:OpticalCoherence}.

The paper is organized as follows. In Sec.~\rsect{operator}, we overview the theory of squeezing and present the solution in an operator form. In the following Sec.~\rsect{array}, we formulate equations for photon numbers in Fock representation, and demonstrate their mapping to classical light propagation in specially designed arrays of optical waveguides. In Sec.~\rsect{bloch}, we discuss the photon state dynamics and the critical transition to Bloch oscillations associated with the variation of the phase mismatch. In Sec.~\rsect{QPM} we analyze the effect of periodic reversal of the sign of nonlinear coefficient, leading to quasi-phase-matching and quasi-Bloch oscillations. Finally, we present conclusions and outlook in Sec.~\rsect{conclusions}.

\section{Two-mode squeezing in nonlinear media with arbitrary phase mismatch} \lsect{operator}

We consider two-mode squeezing in the traveling-wave configuration. The theory of such process is well developed~\cite{*[{}] [{, Sec. 22.4.}] Mandel:1995:OpticalCoherence, *[{}] [{, Sec. IV.A.}] Kolobov:1999-1539:RMP}. In this section, we use the established approach to derive the photon state evolution in the general case of varying phase mismatch between the pump and the squeezed photons. Such analysis is essential for our study, as we later demonstrate the critical transition of light dynamics associated with the variation of the phase mismatch. Specifically, we consider the Hamiltonian in the undepleted pump approximation,
\begin{equation} \leqt{H}
  \hat{H} = \beta_s(z) a_s^\dag a_s + \beta_i(z) a_i^\dag a_i
        + \gamma(z) a_s^\dag a_i^\dag + \gamma^\ast(z) a_s a_i ,
\end{equation}
where $z$ is the propagation direction, $a_{s,i}^\dag$ and $a_{s,i}$ are the creation and annihilation operators for the signal and idler photons, $\gamma$ is proportional to the pump amplitude and quadratic nonlinear susceptibility, and $\beta_{s,i}$ characterize the phase mismatch between the signal, idler, and the pump. Importantly, we take into account arbitrary dependance of the mismatch and nonlinearity coefficients on the propagation coordinate. Such dependence can be realized experimentally, for example through domain reversal~\cite{Fejer:1992-2631:IQE} and waveguide tapering.

\subsection{General solution for the operator evolution}

Following~\cite{*[{}] [{, Sec. 22.4.}] Mandel:1995:OpticalCoherence, *[{}] [{, Sec. IV.A.}] Kolobov:1999-1539:RMP} we obtain Heisenberg equations for the evolution of the operators,
\begin{equation} \leqt{dOperatorsH}
  i \frac{d \hat{a}_s}{d z} = \left[ \hat{a}_s, \hat{H} \right], \quad
  i \frac{d \hat{a}_i}{d z} = \left[ \hat{a}_i, \hat{H} \right].
\end{equation}
Substituting Eq.~\reqt{H} into Eq.~\reqt{dOperatorsH}, we obtain
\begin{equation} \leqt{dOperators}
 \begin{split}
  i \frac{d \hat{a}_s}{d z} = \beta_s(z) \hat{a}_s + \gamma(z) \hat{a}_i^\dag , \\
  i \frac{d \hat{a}_i}{d z} = \beta_i(z) \hat{a}_i + \gamma(z) \hat{a}_s^\dag .
 \end{split}
\end{equation}
We seek solution of Eq.~\reqt{dOperators} in the form
\begin{equation} \leqt{OperatorsModesUV}
 \begin{split}
  \hat{a}_s(z) &= \left[ U(z) \hat{a}_s(0) + V(z) \hat{a}_i^\dag(0) \right] \exp[ - i \tilde{\delta}(z) ] , \\
  \hat{a}_i(z) &= \left[ U(z) \hat{a}_i(0) + V(z) \hat{a}_s^\dag(0) \right] \exp[ i \tilde{\delta}(z) ] ,
 \end{split}
\end{equation}
with the initial conditions
\begin{equation} \leqt{UVinitial}
  U(0) = 1, \quad V(0) = 0 .
\end{equation}
After substituting Eq.~\reqt{OperatorsModesUV} into Eq.~\reqt{dOperators}, we get
\begin{eqnarray}
     i \frac{d U}{d z} &=& \beta(z) U + \gamma(z) V^\ast ,  \leqt{Ueq} \\
     i \frac{d V}{d z} &=& \beta(z) V + \gamma(z) U^\ast ,  \leqt{Veq}
\end{eqnarray}
and
\begin{equation} \leqt{tdelta}
  \tilde{\delta} = \int_0^z \delta(\xi) d \xi,
\end{equation}
where
\begin{equation} \leqt{beta-delta}
 \beta(z) = \frac{\beta_s(z) + \beta_i(z)}{2}, \quad \delta(z) = \frac{\beta_s(z) - \beta_i(z)}{2} .
\end{equation}

It can be checked that $(|U|^2 - |V|^2)$ is a conserved quantity of Eqs.~\reqt{Ueq},\reqt{Veq}. Then, we can use the representation
\begin{equation} \leqt{UVtransform}
  U = U_\alpha \exp( i \varphi) , \quad V = V_\alpha \exp(i \varphi) ,
\end{equation}
where
\begin{equation} \leqt{UValpha}
   U_\alpha = \cosh( |\alpha| ), \quad
   V_\alpha = -\exp\left[ i\, {\rm arg}(\alpha) \right] \sinh( |\alpha| ) ,
\end{equation}
and
\begin{equation} \leqt{alpha}
   \begin{split}
      \alpha &= \tanh^{-1}\left( \left|\frac{V}{U}\right| \right)
                 \exp\left[ i\, {\rm arg}\left( - \frac{V}{U} \right) \right], \\
      \varphi &= {\rm arg}\left( U \right).
   \end{split}
\end{equation}
The operator solution can then be written as:
\begin{equation} \leqt{OperatorsMatrix}
  \left(  \begin{array}{l} \hat{a}_s(z) \\ \hat{a}_i^\dag(z) \end{array} \right) =
  \left(  \begin{array}{ll} e^{i \varphi - i \tilde{\delta} } & 0\\ 0 & e^{- i \varphi - i \tilde{\delta} } \end{array} \right)
  \left(  \begin{array}{ll} U_\alpha & V_\alpha\\ U_\alpha^\ast & V_\alpha^\ast \end{array} \right)
  \left(  \begin{array}{l} \hat{a}_s(0) \\ \hat{a}_i^\dag(0) \end{array} \right)
\end{equation}

\subsection{Solution in terms of standard two-mode squeezing operator}

Let us for a moment consider the case of zero detuning ($\beta_s = \beta_i = 0$) and constant nonlinearity ($\gamma = {\rm const}$). Then $\varphi = \tilde{\delta} = \beta = 0$ and the operator solution is given by Eq.~\reqt{OperatorsMatrix} with $\{U,V\} \equiv \{U,V\}_\alpha$ and $\alpha = i \gamma z$. On the other hand, the operator solution can be equivalently written in a general form as~\cite{*[{}] [{, Sec. 7.7.}]  Gerry:2004:OpticalCoherence}
\begin{equation} \leqt{OperatorsSqueezingDegenerate}
  \left(  \begin{array}{l} \hat{a}_s(z) \\ \hat{a}_i^\dag(z) \end{array} \right) =
     \hat{S}_\alpha^\dag
      \left(  \begin{array}{l} \hat{a}_s(0) \\ \hat{a}_i^\dag(0) \end{array} \right)
     \hat{S}_\alpha
\end{equation}
where the two-mode squeezing or two-photon displacement operator is
\begin{equation} \leqt{S_alpha}
  \hat{S}_\alpha = \exp( -i \hat{H}_\alpha ) = \exp( \alpha^\ast a_s a_i - \alpha a_s^\dag a_i^\dag )
\end{equation}

Now considering the case of arbitrary detuning and nonlinearity parameters, and taking into account the equivalent representations discussed above, we can write the operator solution~\reqt{OperatorsMatrix} as
\begin{equation} \leqt{OperatorsSqueezing}
 \begin{split}
  \left(  \begin{array}{l} \hat{a}_s(z) \\ \hat{a}_i^\dag(z) \end{array} \right) = &
   \exp\left( i \hat{H}^{(s)} \right)
   \exp\left( i \hat{H}^{(i)} \right)
       \hat{S}_\alpha^\dag
          \left(  \begin{array}{l} \hat{a}_s(0) \\ \hat{a}_i^\dag(0) \end{array} \right) \\
      & \cdot \hat{S}_\alpha
   \exp\left( - i \hat{H}^{(i)} \right)
   \exp\left( - i \hat{H}^{(s)} \right)
 \end{split}
\end{equation}
where
\begin{equation} \leqt{H-si}
  \hat{H}^{(s)} = (\tilde{\delta} - \varphi) a_s^\dag a_s, \quad \hat{H}^{(i)} = (-\tilde{\delta} - \varphi) a_i^\dag a_i
\end{equation}
Accordingly, in an interaction picture solution can be represented as
\begin{equation} \leqt{Psi_alpha_detuning}
   | \Psi(z) \rangle =  \exp\left( - i \hat{H}^{(i)} \right) \exp\left( - i \hat{H}^{(s)} \right) \hat{S}_\alpha | \Psi(0) \rangle
\end{equation}

We can now calculate the average photon number evolution. Note that the terms $\exp( -i \hat{H}^{(i)} )$ and $\exp( -i \hat{H}^{(s)})$ can be neglected in this calculation, since these operators preserve the photon numbers:
\begin{equation} \leqt{PhotonNumber}
    \begin{split}
      \langle n_s(z) \rangle &= \langle \hat{a}_s^\dag(z) \hat{a}_s(z) \rangle \\
      &= \langle \Psi(0) | \hat{S}_\alpha^\dag  \hat{a}_s^\dag(0) \hat{a}_s(0) \hat{S}_\alpha | \Psi(0) \rangle \\
      &= \langle n_s(0) \rangle + |V|^2 \left[1 + \langle n_s(0) \rangle  + \langle n_i(0) \rangle \right] \\
         &+ \langle \Psi(0) | \left[ U^\ast V \hat{a}_s^\dag(0) \hat{a}_i^\dag(0) + U V^\ast \hat{a}_s(0) \hat{a}_i(0) \right] | \Psi(0) \rangle \\
    \end{split}
\end{equation}

If the input state has a fixed photon number, i.e. $| \Psi(0) \rangle = | n_s^{(0)}, n_i^{(0)} \rangle$, then
\begin{equation} \leqt{PhotonNumberFixedInput}
    \begin{split}
        \langle n_s(z) \rangle &= n_s^{(0)} + |V|^2 \left[ 1 + n_s^{(0)} + n_i^{(0)} \right], \\
        \langle n_i(z) \rangle &= n_i^{(0)} + |V|^2 \left[ 1 + n_s^{(0)} + n_i^{(0)} \right] .
    \end{split}
\end{equation}

\section{Waveguide arrays design for simulating Fock state evolution} \lsect{array}

\subsection{Coupled-mode equations for the photon numbers}

Two-mode squeezed state in the number (Fock) basis can be written as~\cite{*[{}] [{, Sec. 7.7.}]  Gerry:2004:OpticalCoherence}:
\begin{equation} \leqt{Psi}
  |\Psi(z)\rangle = \sum_{n_s=0}^{+\infty} \sum_{n_i=0}^{+\infty} \psi_{n_s,n_i}(z) |n_s,n_i \rangle .
\end{equation}
Equations for the evolution of the state vector are
\begin{equation} \leqt{dPsiH}
  i \frac{d \Psi(z)}{d z} = \hat{H} \Psi(z).
\end{equation}

Since the photons are generated in pairs, only the states with
\begin{equation} \leqt{N}
   n_i-n_s = N = {\rm const}
\end{equation}
will be coupled. Since the signal and idler equations are symmetric, with no loss of generality we consider $N \ge 0$, and obtain the coupled-mode equations with modulated coupling for $\psi_n = \psi_{n,n+N}$,
\begin{eqnarray}
    i \frac{d \psi_0}{d z} &=& c_1^\ast \psi_1 + N \beta_i \psi_0, \leqt{dpsi-0} \\
    i \frac{d \psi_n}{d z} &=& c_n \psi_{n-1} + c_{n+1}^\ast \psi_{n+1}  \nonumber \\
                              &&+ \left[n \beta_s + (n+N) \beta_i \right] \psi_n,  \leqt{dpsi-n}
\end{eqnarray}
where $n \ge 1$ and $c_n$ are the modulated coupling coefficients:
\begin{equation} \leqt{Cdifferent}
  c_n(z) = \sqrt{ n (n+N) } \gamma(z) ,
\end{equation}

We now simplify the coupled equations. First, we make a transformation of the propagation coordinate,
\begin{equation} \leqt{Zgamma}
   Z(z) = \gamma_0^{-1} \int_0^z |\gamma(\xi)| d \xi ,
\end{equation}
where $\gamma_0$ is a positive scaling coefficient, and obtain
\begin{eqnarray}
    i \frac{d \psi_0}{d Z} &=& \tilde{C}_1^\ast \psi_1 + N \beta_i  (\gamma_0 / |\gamma(Z)|) \psi_0, \leqt{dpsiZ-0} \\
    i \frac{d \psi_n}{d Z} &=& \tilde{C}_n \psi_{n-1} + C_{n+1}^\ast \psi_{n+1}    \nonumber \\
                           &&+ \left[n \beta_s + (n+N) \beta_i \right] (\gamma_0 / |\tilde{\gamma}(Z)|) \psi_n,  \leqt{dpsiZ-n}
\end{eqnarray}
where
\begin{equation} \leqt{Ctilde}
  C_n(z) = \sqrt{ n (n+N) } \gamma_0 \exp\left[i\, q(Z)\right],
  \quad \tilde{\gamma}(Z) = \gamma(z),
\end{equation}
and
\begin{equation} \leqt{qn}
   q(Z) = {\rm arg}\left[ \tilde{\gamma}(Z) \right]
\end{equation}
Next, we make a substitution
\begin{equation} \leqt{psiTilde}
   \psi_n(Z) = E_n(Z) \exp\left( i q n - i N \int_0^{Z} \beta_i(\xi) \frac{\gamma_0}{|\tilde{\gamma}(\xi)|} d \xi \right),
\end{equation}
and obtain equations with positive $Z$-independent coupling coefficients
\begin{eqnarray}
    i \frac{d E_0}{d Z} &=& |C_1| E_1 , \leqt{dpsit-0} \\
    i \frac{d E_n}{d Z} &=& |C_n| E_{n-1} + |C_{n+1}| E_{n+1}  + \rho(Z) n E_n,  \leqt{dpsit-n}
\end{eqnarray}
where
\begin{equation} \leqt{rho}
    \rho(Z) =  \left(2 \beta(Z) - \frac{d q}{d Z} \right) \frac{\gamma_0 }{  |\tilde{\gamma}(Z)|} .
\end{equation}

To perform an optical simulation of Fock-space dynamics, one could use straight arrays of coupled waveguides, with parameters designed to match Eqs.~\reqt{dpsit-0},\reqt{dpsit-n}. This would be similar to Ref.~\cite{Keil:2012-3801:OL}, but with different waveguide coupling.

\pict{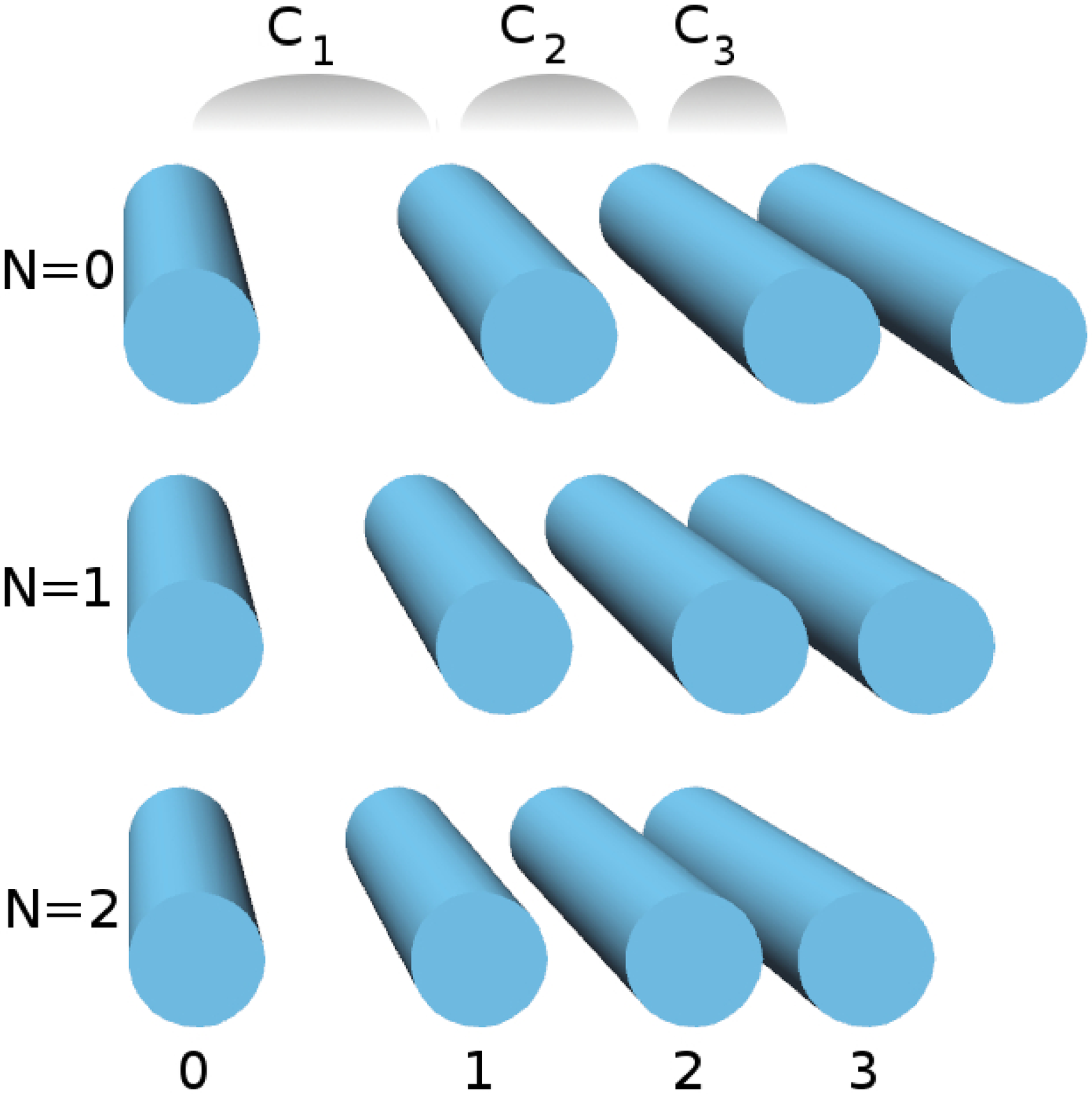}{arrays}{
(Color online)
Schematics of waveguide arrays with varying waveguide separation to achieve the coupling $|C_n|$ according to Eq.~\reqt{Ctilde}. The coupling and array design depends on the difference between the signal and idler photon numbers: (a)~$N=0$, (b)~$N=1$, (c)~$N=2$.
}

\section{Bloch oscillations of squeezed light} \lsect{bloch}

\subsection{Transition from non-oscillatory regime to Bloch oscillations}

We now consider the constant ($z$-independent) coefficients $\beta_{s,i}$ and $\gamma$.
Then, the solution of Eqs.~\reqt{Ueq},\reqt{Veq} for the operator amplitudes can be written explicitly as
\begin{equation} \leqt{UV}
 \begin{split}
     U &= \cosh( b z ) - \frac{i \beta}{b} \sinh( b z )  , \\
     V &= - i \frac{\gamma}{b} \sinh( b z ) .
 \end{split}
\end{equation}
where
\begin{equation} \leqt{betaCr}
   b = \sqrt{ \beta_{\rm cr}^2 - \beta^2}, \quad \beta_{\rm cr} = |\gamma|.
\end{equation}
We then use Eq.~\reqt{PhotonNumberFixedInput} to find the average photon number evolution for an input state with a fixed photon number, $| \Psi(0) \rangle = | n_s^{(0)}, n_i^{(0)} \rangle$, as
\begin{equation} \leqt{PhotonNumberFixedInputBO}
    \begin{split}
        \langle n_s(z) \rangle &= \langle n_i(z) \rangle - N \\
                              &= n_s^{(0)} + \frac{|\gamma|^2}{|b|^2}
                                  \left|\sinh( b z )\right|^2 \left[ 1 + N + 2 n_s^{(0)} \right], \\
    \end{split}
\end{equation}
where according to Eq.~\reqt{N}, $N = n_i^{(0)} - n_s^{(0)}$.

We see that the type of evolution fundamentally changes when the detuning crosses the critical value $\beta_{\rm cr}$.

For detunings smaller than the critical value, $|\beta| < \beta_{\rm cr}$, $b$ is real, and at large $z$ the values of $U$ and $V$ will rapidly grow corresponding to quickly increasing photon numbers according to Eq.~\reqt{PhotonNumber}.

For detunings larger than the critical value, $|\beta| > \beta_{\rm cr}$, $b$ is imaginary, and solution becomes oscillating with the period
\begin{equation} \leqt{BlochPeriod}
  \Delta z = \frac{\pi}{|b|} = \frac{\pi}{ \sqrt{ \beta^2 - \beta_{\rm cr}^2 }} .
\end{equation}
At distances $z = m \Delta z$, where $m=1,2,3,4,\ldots$, the solution returns to the input state.
Note that such dynamics will occur for any input, in lattices corresponding to any value of $N$.
Considering the realization in photonic lattices,
we can call this regime {\em Bloch oscillations of squeezed light} as a generalization of spatial Bloch oscillations in waveguide arrays~\cite{Peschel:1998-1701:OL, Pertsch:1999-4752:PRL, Morandotti:1999-4756:PRL} or {\em squeezed quantum bouncing ball} as a generalization of quantum bouncing ball in photonic lattices~\cite{Longhi:2008-35802:PRA, DellaValle:2009-180402:PRL}.

In the special case of equal signal and idler photon numbers, when $N=0$, the coupling and detuning coefficients in Eqs.~\reqt{dpsit-0},\reqt{dpsit-n} directly correspond to the previously studied lattice with linearly increasing hopping rate and on-site potential~\cite{Longhi:2009-33106:PRB}. In Ref.~\cite{Longhi:2009-33106:PRB}, a transition between Bloch oscillations and non-periodic dynamics was first identified, which agrees with our findings for a more general coupling dependence corresponding to arbitrary $N$. Actually, we find that the lattice modulation parameters in Eqs.~\reqt{dpsit-0},\reqt{dpsit-n} belong to a previously identified class of lattices with commensurate energy levels~\cite{Longhi:2010-41106:PRB}, where Bloch oscillations were predicted. Specifically, our model corresponds, up to a gauge transformation, to Eqs.~(1),(2) in Ref.~\cite{Longhi:2010-41106:PRB} with $F_1(n) = \gamma_0^{1/2} q (n+N+1)$ and $F_2(n) = \gamma_0^{1/2} q^{-1} n$, where $q$ is defined to satisfy the relation $(q+q^{-1}) = - \rho \gamma_0^{-1/2}$.

There were also recent studies of Bloch-like oscillations in Jaynes-Cummings~\cite{Longhi:2011-3407:OL} and Glauber-Fock photonic lattices~\cite{Perez-Leija:2012-13848:PRA, Keil:2012-3801:OL}, where the oscillation period was found to be inversely proportional to the linear detuning between the waveguides ($\beta$ in our notations).
These results can be obtained as a limiting case of our study by taking $N \rightarrow \infty$ and $\gamma = \gamma_0 \rightarrow \gamma_{\infty} / \sqrt{N}$. In this case, $C_n \rightarrow \gamma_{\infty} \sqrt{n}$ and $\Delta z = 2 \pi / |\beta|$, as for the Glauber-Fock lattice considered previously~\cite{Perez-Leija:2012-13848:PRA, Keil:2012-3801:OL}.

\subsection{Evolution of the input vacuum state}

\pict{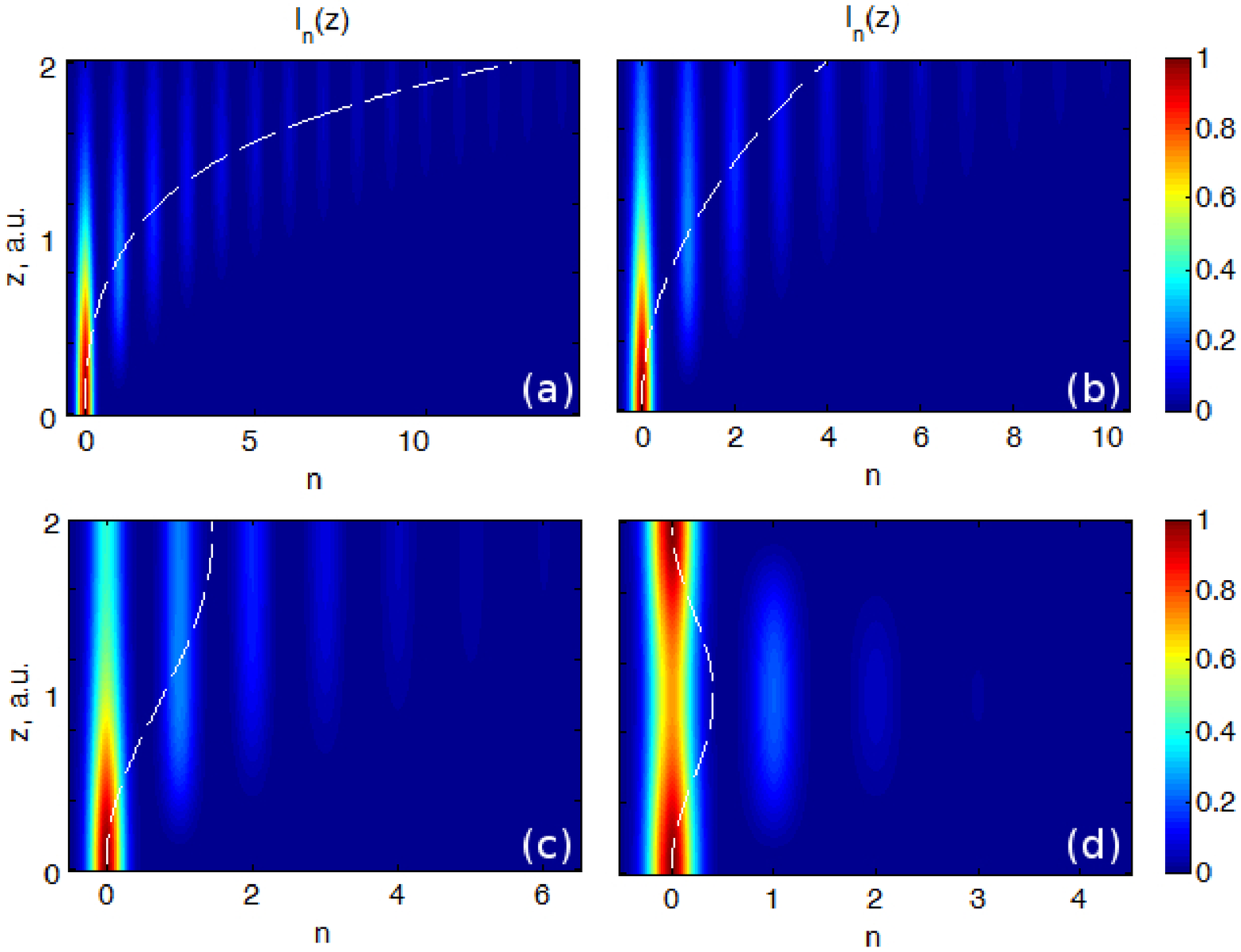}{vacuum}{
(Color online)
Evolution of light intensity in waveguide arrays simulating the photon number distribution in squeezed vacuum states, for different values of the phase mismatch $\beta$: (a)~$0$, (b)~$1$, (c)~$1.3$, and (d)~$1.86$.
Dashed lines show the average signal photon number.
For all the plots, $\gamma=1$ and $N=0$.
}

We first illustrate the general results for the squeezed vacuum state. The vacuum state at the input corresponds to the initial conditions
\begin{equation} \leqt{Input}
  N=0, \quad \psi_1(z=0) = 1, \quad \psi_{n > 1}(z=0) = 0 .
\end{equation}
Combining the expression for the squeezed vacuum state~\cite{*[{}] [{, Sec. 7.7.}] Gerry:2004:OpticalCoherence} with Eq.~\reqt{Psi_alpha_detuning}, we obtain
\begin{equation} \leqt{SV}
  \psi_n(z) = e^{2 i \varphi n} \frac{1}{\cosh(|\alpha|)} (-1)^n e^{i n \theta} \left[ \tanh(|\alpha|) \right]^n
\end{equation}
For convenience, we can choose the scaling coefficient $\gamma_0 = |\gamma|$, such that $Z \equiv z$, and then the normalized mode amplitudes are found as
\begin{equation} \leqt{SVTsimplify}
    \begin{split}
      E_n(Z) = & \left[ \cosh^2( b Z ) + \beta^2 |b^{-1} \sinh( b Z )|^2\right]^{-1/2} \\
          & \cdot \left[ \frac{- i |\gamma| b^{-1}\sinh( b Z)}{\cosh( b Z ) + i \beta b^{-1} \sinh( b Z )} \right]^n
    \end{split}
\end{equation}

We plot the characteristic dependencies of the mode intensities in Figs.~\rpict{vacuum}(a-d) for different values of the phase mismatch.
In the phase-matched regime, the photon number increases rapidly, as marked with the dashed line in Fig.~\rpict{vacuum}(a). For phase mismatch chosen at the threshold, $\beta - \beta_{cr}$, the photon number also grows continuously, but at a slower rate, see Fig.~\rpict{vacuum}(b). However, as the mismatch is increased above the threshold value, the photon number growth becomes bounded and the photon distribution periodically returns to the input state, see Figs.~\rpict{vacuum}(c,d). In particular, the oscillation period is $\Delta z = 2$ for the mismatch parameter in Fig.~\rpict{vacuum}(d).

\subsection{Synchronous Bloch oscillations of squeezed quantum bouncing ball}

\pict{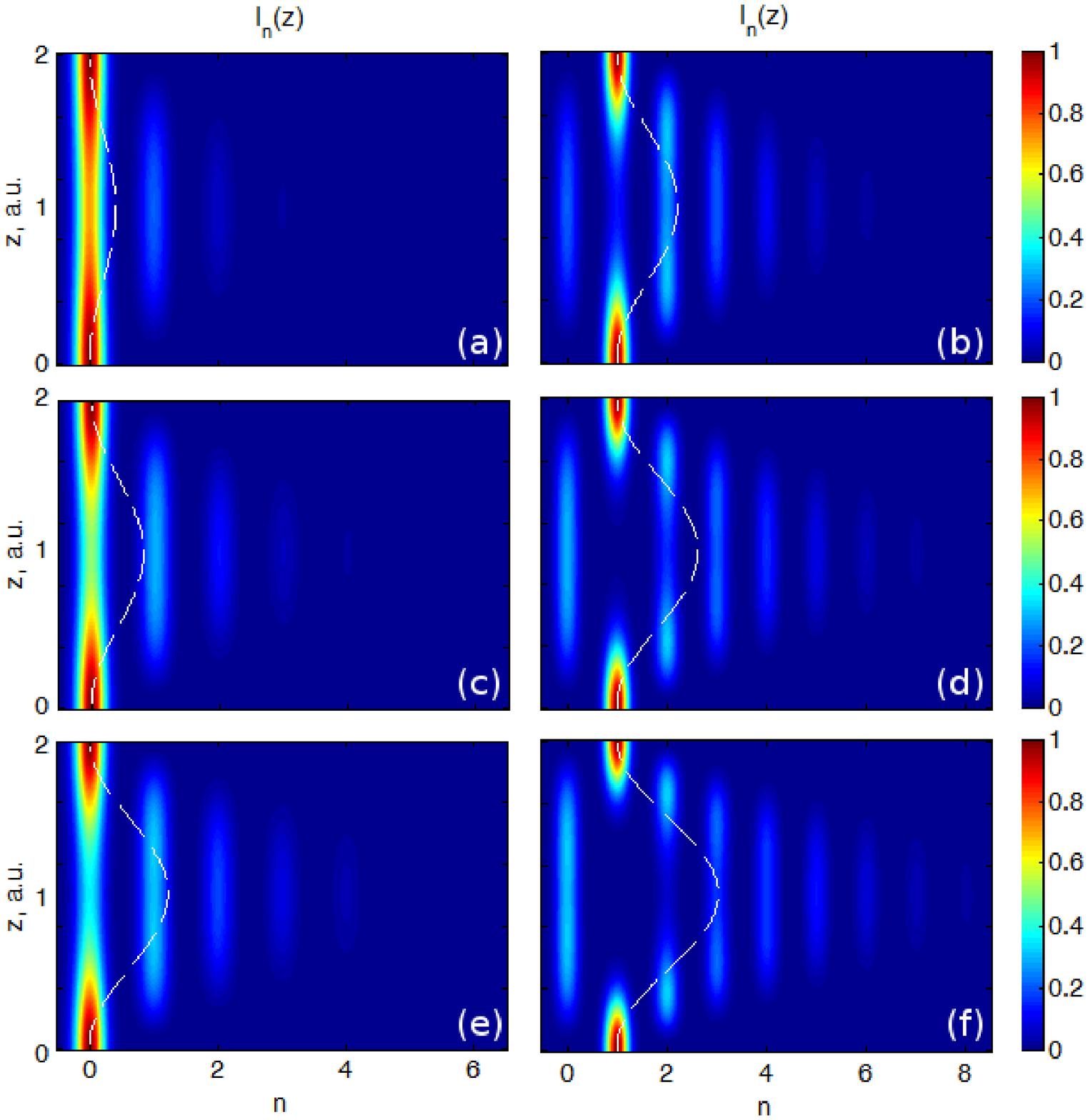}{bloch}{
(Color online)
Synchronous Bloch oscillations in waveguide arrays corresponding to various input conditions: (a,c,e)~input at the waveguide $n=0$ corresponding to $n_s^{(0)}=0$ and (b,d,f)~input at the waveguide corresponding to $n_s^{(0)}=1$.
Dashed lines show the average signal photon number.
The photon population differences between the signal and idler photons are (a,b)~$N=0$, (c,d)~$N=1$, (c,d)~$N=2$, and these are simulated using arrays with different couplings as illustrated in Fig.~\rpict{arrays}.
For all the plots, $\gamma=1$ and $\beta = 1.86$.
}

We now illustrate that for arbitrary input, the Bloch oscillations occur synchronously, with the same oscillation period according to Eq.~\reqt{BlochPeriod}. We fix the values of the nonlinear and phase mismatch coefficients, and perform numerical simulations corresponding to different initial conditions in Fig.~\rpict{bloch}. The three rows in this figure show the evolution for the photon number differences between the signal and idler modes $N=0,1,2$, and these correspond to accordingly different configurations of waveguide arrays as illustrated in Fig.~\rpict{arrays}. The left and right columns in Fig.~\rpict{bloch} correspond to different excitation condition of each array, where the light is coupled to the first or second waveguide at the input, respectively. Such coupling corresponds to the input states with zero ($n_s^{(0)}=0$) or one ($n_s^{(0)}=1$) photon in the signal mode at the input. We observe that the photon number distributions are strongly dependent on the input number of photons in the signal and idler modes. Nevertheless, in agreement with the analytical predictions, the full revival of an input state is observed after on Bloch oscillation period.

Such features of Bloch oscillations with the periodic revivals indicate that the spectrum of eigenmodes in the lattices is equidistant, forming a semi-infinite Wannier-Stark ladder with the spacing between the levels of $2 \pi / \Delta z$. Moreover, this spectrum is the same for different arrays corresponding to various values of $N$. Therefore, we have identified an infinite family of lattices with different non-uniform coupling parameters, but the same equidistant spectra. This can offer new opportunities for optical beam manipulation.

\section{Periodic reconstruction and squeezing in quasi-phase-matched structures} \lsect{QPM}

In media with quadratic nonlinearity, the efficiency of interactions can be increased under the presence of phase mismatch through the quasi-phase-matching (QPM) technique~\cite{Fejer:1992-2631:IQE}. With this approach, the sign of the quadratic nonlinear coefficient is reversed along the propagation direction. We analyze the effect of such modulation of the simulation of squeezing dynamics in waveguide arrays, considering a periodic modulation of the sign of the nonlinear coefficient. Specifically, we take $\gamma(x) = \gamma_0 > 0$ for $0 \le x < L/2$ and $\gamma(x) = - \gamma(x-L/2)$ for $x \ge L/2$, where $L$ is the modulation period.
In terms of the coupled-mode Eqs.~\reqt{dpsit-0},\reqt{dpsit-0}, each the sign reversals correspond to $\rho(Z) = \pi \sum_m \delta(Z - m L/2)$, where $\delta(Z)$ is Dirac delta function. Accordingly, at these positions the phase of the wave function is abruptly changed as $E_n(m L/2 + \xi) = (-1)^n E_n(m L/2 - \xi)$ for $\xi \rightarrow +0$. Such phase modulation was suggested previously for the realization of image reconstruction~\cite{Longhi:2008-473:OL, Szameit:2008-181109:APL}, however we find that in our system the possibility of input state reconstruction strongly depends on the phase mismatch.

\pict{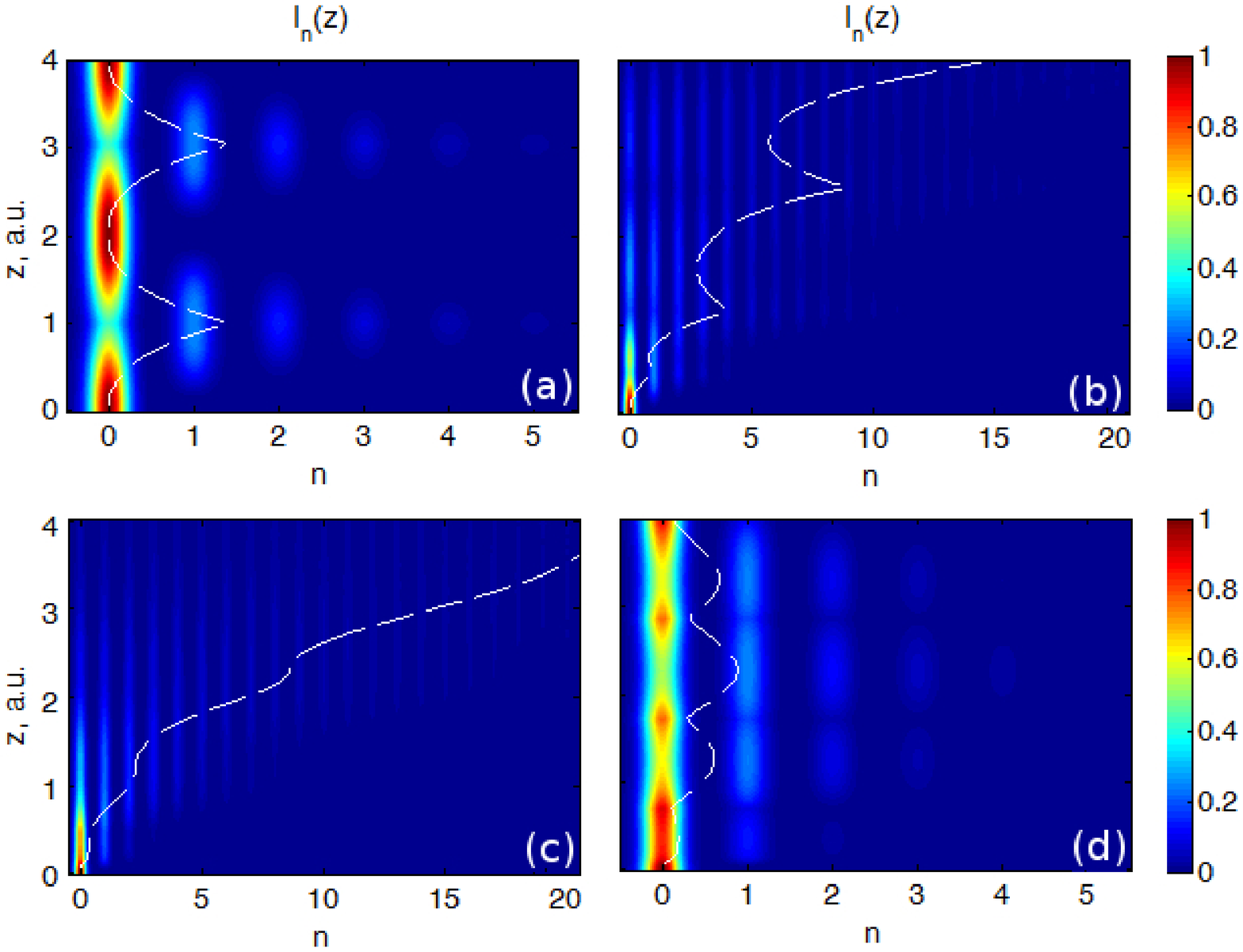}{QPM}{
(Color online)
Effect of periodic sign reversal of the nonlinear coefficient with the period $L=2$ for different values of the phase mismatch $\beta$: (a)~$0$, (b)~$1$, (c)~$1.86$, and (d)~$2.5$.
Dashed lines show the average signal photon number.
For all the plots, $|\gamma(x)|=1$ and $N=0$.
}

We present the simulation results in Figs.~\rpict{QPM}(a-d) for the evolution of an input vacuum state with different values of the phase mismatch. In the case of exact phase matching ($\beta=0$), the reversal of the sign of nonlinear coefficient effectively reverses the propagation dynamics, such that after each modulation period the input state is restored, similar to Refs.~\cite{Longhi:2008-473:OL, Szameit:2008-181109:APL}. As a phase mismatch is introduced, the reversals are suppressed, see Figs.~\rpict{QPM}(b,c). This agrees with the principles of QPM, that the effective phase mismatch is modified by the reciprocal wavevector $2 \pi / L$. The fastest increase of the photon number is visible in Fig.~\rpict{QPM}(c), when the mismatch ($\beta = 1.86$) is chosen such that the Bloch oscillation period in the absence of QPM [Fig.~\rpict{vacuum}(d)] exactly matches the QPM period. For particular values of mismatches, the revival of the input state can occur as shown in Fig.~\rpict{QPM}(d). The latter case resembles the regime of quasi-Bloch oscillations and dynamic localization~\cite{Wan:2004-125311:PRB}.

\section{Conclusions} \lsect{conclusions}

In conclusion, we have demonstrated that arrays of optical waveguides with specially modulated coupling can directly model the photon number evolution in the process of quantum two-mode squeezing, in the general case of coordinate-dependant phase mismatch and nonlinearity coefficients. We identified a phase-mismatch dependent transition between a gradual beam diffraction corresponding to an increase of the squeezed photon numbers, and Bloch oscillations with the periodic revival of an arbitrary input state.
Additionally, we have illustrated a different regime of periodic revivals associated with the periodic reversal of the effective nonlinear coefficient, resembling the regimes of input reconstruction and quasi-Bloch oscillations.

We anticipate that our results on the families of waveguide arrays with modulated parameters may stimulate
further studies towards novel opportunities of employing such lattices for manipulation of optical beams including all-optical switching of intense beams~\cite{Lederer:2008-1:PRP, Garanovich:2012-1:PRP} and control of spatial entanglement through quantum walks with correlated photons~\cite{Peruzzo:2010-1500:SCI}.

The work was supported by the Australian Research Council, including Discovery Project DP130100135 and Future Fellowship FT100100160.

\end{document}